\documentclass[11pt]{article}  
\usepackage{graphicx}
\usepackage{natbib}
\usepackage{graphicx}
\usepackage[latin1]{inputenc}
\usepackage{latexsym}

\begin{document}
\title{The EUV Sun as the superposition of elementary Suns}

\author{P.-O. Amblard$^1$, S. Moussaoui$^2$, T. Dudok de Wit$^3$, J. Aboudarham$^4$, \\ M. Kretzschmar$^5$, J. Lilensten$^6$, F. Auch\`ere$^7$}

\date{\small{
$^1$ GIPSA-lab, DIS (UMR CNRS 5216, INPGrenoble)
 BP46, 38402 Saint-Martin d'H\`eres, France, \\
$^2$ IRCCYN (UMR CNRS 6597,  ECNantes), 1 rue de la No\'e, BP92101, 44321 Nantes cedex 3, France\\   
$^3$ LPCE (UMR 6115 CNRS-Universit\'e d'Orl\'eans),
3A avenue de la Recherche Scientifique, 45071 Orl\'eans, France\\
$^4$ LESIA (UMR CNRS 8109 , Observatoire de Paris), 5 Place Jules Janssen, 92195 Meudon, France\\
$^5$ LPG (UMR 5109 CNRS-Universit\'e Joseph Fourier), B\^atiment de Physique D, BP 53, 38041 Saint Martin d'H\`eres Cedex, France\\
$^6$ IAS (UMR 8617 CNRS-Universit\'e Paris-Sud), 91045 Orsay, France
}}

   \maketitle

\begin{center}
\textbf{Abstract}
\end{center}

Many studies assume that the solar irradiance in the EUV can be decomposed into different contributions, which makes the modelling of the spectral variability considerably easier. We consider a different approach, in which these contributions are not imposed a priori but are effectively and robustly inferred from spectral irradiance measurements. This is a source separation problem with a positivity constraint, for which we use a Bayesian solution. Using five years of daily EUV spectra recorded by the TIMED/SEE satellite, we show that the spectral irradiance can be decomposed into three elementary spectra. Our results suggest that they describe different layers of the solar atmosphere rather than specific regions. The temporal variability of these spectra is discussed. \\

\noindent Published in : Astronomy and Astrophysics, Volume 487, Issue 2 (2008), pp.L13-L16

\sloppy

\section{Motivation}

The solar extreme ultraviolet (EUV) irradiance is the primary energy input for the diurnal ionosphere and one of the key parameters for space weather. Any variation in the incoming EUV flux at the top of Earth's atmosphere modifies the state of the thermosphere/ionosphere system and can affect human activities such as radio telecommunication and orbitography (through satellite drag). The knowledge of the spectral irradiance in real time is compulsory for mitigating its potentially harmful effects.

The EUV flux, however, can only be measured from space. Several approaches have been developed to overcome this dependence on space-borne instruments. Historically, and driven in part by the lack of measurements, indices such as the radio flux at 10.7 cm \citep{richards94,tobiska00} have been used as proxies for the solar EUV flux. For a long time, these indices have provided ionospheric physicists with very useful inputs for their models. These indices, however, also have intrinsic limits because of the difference between the physical processes that give rise to them and to the EUV emission. Furthermore, the observation of the solar disk in the EUV, notably through the SoHO satellite, has revealed tremendous heterogeneity and dynamics in this spectral range. It then comes as no surprise that the whole EUV spectrum variability can hardly be reproduced with a single proxy \citep{floyd05,ddw08}.

A different approach consists in decomposing the solar spectral irradiance into the sum of contributions that come from different regions, each of which has a characteristic spectrum. For example, \citet{vernazza78}, using SKYLAB data, and later \citet{curdt01}, using SOHO/SUMER data, empirically decomposed the Sun into three regions (quiet Sun, coronal holes, and active regions) and associated a  typical spectrum to each of them. Similar ideas were put forward by \citet{lean82} and \citet{woods00}. This has led \cite{warren01} to model the solar EUV irradiance as a linear combination of three spectra that are again associated with the quiet Sun, coronal holes, and active regions. \citet{kretzschmar04} and \citet{warren05} have pursued these studies. Good agreement has been found with other models and measurements \citep{woods05}. A similar strategy has been used for the near-UV and visible range, where the solar surface has been decomposed into photospheric features such as sunspots and faculae  \citep{fontenla05,wenzler06}.  

All these studies, however, rely on the rather subjective choice of solar regions and on the assumption that these may be associated with characteristic (or elementary) spectra. This strong constraint has always been justified through empirical arguments. A first problem here is to determine the number of elementary spectra. A second problem is to define the solar regions and the resolution needed to resolve them. One may, for example, wonder how small the solar features should be to properly explain the variability of the whole disk. This is a kind of endless problem, since the better the resolution, the finer the structure and the stronger the dynamics. 

To bypass these problems, we follow a  novel approach. Instead of starting from a predefined set of solar regions that are guessed from empirical knowledge, we use a statistical method to determine if the solar EUV spectrum can be decomposed at all, and to extract its different components. A major difference with respect to previous approaches is the identification of elementary spectra that are based on only the statistical properties of the solar spectral dynamics, without any a priori on the number or on the shape of these spectra. In this sense, the method is less biased and we are more likely to discover new and unsuspected aspects of the solar variability in the EUV.

The method we use is based on a recent and powerful mathematical concept called Bayesian positive source separation (BPSS), allows us here to decompose the solar EUV spectral variability into a linear superposition of contributions. The motivation of this letter is twofold. First, we show that three elementary spectra are sufficient for reproducing the salient features of the EUV spectral variability. Second, we show that these elementary spectra, which are determined by statistical means alone, actually have a physical interpretation. Our results suggest that they describe different volumes of the solar atmosphere rather than specific regions.

%%%%%%%%%%%%%%%%%%%%%%%%%%%%%%%%%%%%%%%%%%%%%%%%%%%%%%%%%%%%%%%%%%%
% \vspace{-.3cm}
\section{Positive source separation and its application to the EUV solar spectrum analysis}

We consider the five years of daily solar spectral irradiance measurements since Feb. 2002 by the Solar EUV Experiment (SEE)  onboard TIMED \citep{woods05}. 
The EUV Grating Spectrograph (EGS), which is part of SEE, measures the spectrum from 25 to 195 {nm} with a 0.4 {nm} spectral resolution. We use  level 2 data (version 9), in which the spectral irradiance is provided from 25 to 195 nm with a 0.1 nm spectral step. Solar flares are excluded from this data set. Some wavelengths are missing around the strong HI Lyman-$\alpha$ line for instrumental reasons. The signal-to-noise ratio gradually decreases in time because of instrument degradation and the declining solar cycle. Our results, however, remain unchanged when the analysis is performed only on the first half of the data set.

The spectral irradiance data from TIMED/SEE are stored in a matrix $\vec{I}(t,\lambda) $ of size $(n_t=2146 ,n_\lambda=1546)$ where $t$ denotes time and $\lambda$ denotes wavelength. Our objective is to decompose each spectrum as a linear combination of $n_e$ elementary spectra or, equivalently, to decompose the matrix $\vec{I}(t,\lambda)$ as a product of two matrices $ \vec{V}(t)$ and $ \vec{S}(\lambda)$, of respective sizes $(n_t,n_e)$ and $ (n_e,n_\lambda)$.  Each line of the matrix $ \vec{S}(\lambda)$   contains an elementary spectrum, and its associated time variability is stored in the corresponding column of the matrix $\vec{V}(t)$. These matrices   are positive, in the sense that all their entries are positive. This problem is known as  {\em  positive matrix factorization } \citep{paatero94} or {\em   positive source separation }  \citep{moussaoui06}.  

The approach   to solve this factorization problem is based on Bayesian estimation theory \citep{gelman03}. We improve the data model by including an additive noise term $\vec{B}(t,\lambda)$, which corresponds to measurement noise and to data modelling errors. In the following, we assume that the entries   $\vec{B}(t,\lambda)$ are independent, zero-mean Gaussian random variables. The positive source separation problem  then consists in  finding the matrices $\vec{V}(t)$ and $ \vec{S}(\lambda)$ from the knowledge  only of the data $\vec{I}(t,\lambda) $ and under the assumption of the modelling equation $\vec{I}(t,\lambda)  =  \vec{V}(t) \times \vec{S}(\lambda)+
\vec{B}(t,\lambda)$. According to the Bayesian paradigm \citep{gelman03}, we assume that $\vec{V}(t)$ and $ \vec{S}(\lambda)$ are random matrices, and the assessment of these matrices is to be understood in a probabilistic sense. In other words, the problem is solved if we know the joint probability distribution of $\vec{V}(t)$ and $ \vec{S}(\lambda)$, given the data $\vec{I}(t,\lambda) $, called the {\it a posteriori} distribution. According to Bayes' theorem,  the {\it a posteriori} distribution writes (omitting $t$ and $\lambda$ for the sake of clarity)
$$
P\left( \vec{S},\vec{V}\big| \vec{I}  \right) = P\left(  \vec{I} \big|  \vec{S},\vec{V}\right)\times P\left( \vec{S},\vec{V}\right) / P(\vec{I}).  
$$
In this equation, $P\left(  \vec{I} \big|  \vec{S},\vec{V}\right)$ is called the likelihood function and will be known if the observation equation or modelling equation is known. Here, since $\vec{I}=\vec{V} \times \vec{S} + \vec{B}$, the likelihood function is simply given by the distribution of the noise term with mean $\vec{V}\times \vec{S}$, or  $P_{\vec{B}}(\vec{I}-\vec{V}\times \vec{S} )$. The second term, $P\left( \vec{S},\vec{V}\right)$, is called the {\it a priori } distribution of the parameters we are looking for. This distribution has to be chosen carefully, according to the {\it a priori } knowlegde we have on the variabilities $\vec{V}$ and the elementary spectra $\vec{S}$. We assume here that these spectra and the variabilities are statistically independent random matrices, so that their joint distribution factorizes into $P(\vec{S})\times P(\vec{V})$. Furthermore, since we are looking for positive quantities, we impose that the distributions are zero for negative values of any of their arguments. Typically, we assume that the entries of the matrices are independent random variables, and identically distributed according to Gamma probability density functions. 

All the assumptions described above allow us to write the {\it a posteriori} distribution $P\left( \vec{S},\vec{V}\big| \vec{I}  \right) $, and knowing this distribution means that all the information contained in the data $\vec{I}$ about the parameters $\vec{S}$ and $\vec{V}$ are known. However, a pragmatic point of view imposes point estimates of the matrices $ \vec{V}(t)$ and $ \vec{S}(\lambda)$. Such estimates are obtained from the {\it a posteriori } distribution by using so-called Bayesian estimators, among which the most famous are the minimum mean square error estimator (MMSE) and the maximum {\it a posteriori} estimator (MAP). The former can be shown to be the {\it a posteriori } mean, {\it i.e. } the mean of the {\it a posteriori } distribution, and the latter is the given by the parameters that maximize the {\it a posteriori } distribution. Here, we choose the MMSE estimator, {e.g.}
$$
\widehat{\vec{V}} = \int \vec{V} P\!\left(\vec{V}\big| \vec{I}  \right) d\vec{V} .
$$

In practice, the {\it a posteriori} distribution lies in a high dimensional space and is mathematically so complex that the Bayesian estimators cannot be evaluated theoretically. Numerical approximations are needed, and for several reasons exposed in \citep{moussaoui06}, a Markov chain Monte-Carlo algorithm is used \citep{gelman03}. Such an algorithm provides a multidimensional Markov chain  $\vec{M}(n), n\geq 0  $ such that the distribution of $\vec{M}(n)$ at iteration $n$ is close to the {\it a posteriori } distribution $P\left( \vec{S},\vec{V}\big| \vec{I}  \right)$. The design of the chain ensures that  these distributions coincide asymptotically ($n\rightarrow +\infty$). Furthermore, if the chain is correctly designed, the coincidence can be obtained rapidly ($n_{coincidence} \sim 10^3$ iterations). The outputs obtained after the coincidence are then used to perform a sample mean. For example, if $\vec{M}(n) = (\vec{V}(n),\vec{S}(n))$, we obtain an approximate MMSE estimator of the variabilites {\it via  } 
$$
\widehat{\vec{V}} \approx N^{-1}  \sum_{n=n_{coincidence}+1}^{n_{coincidence}+N} \vec{V}(n). 
$$
This algorithm is explained in \citet{moussaoui06}, and its source code is available on request.

%%%%%%%%%%%%%%%%%%%%%%%%%%%%%%%%%%%%%%%%%%%%%%%%%%%
% \vspace{-.3cm}
\section{Solar EUV irradiance decomposition results}
The first key question is how many elementary spectra are needed to properly reproduce the solar spectral variability. This can be answered in two different ways. First we decompose the spectral irradiance into $n_e$ elementary spectra, then compute the normalized difference between the measured and the reconstructed spectra $e(t,\lambda) = \left(   I(t,\lambda) - \sum_{i=1}^{n_e} V_i(t) S_i(\lambda)\right) \big/ I(t,\lambda) $ and subsequently consider the normalised mean square error $J = \langle e^2(t,\lambda) \rangle_{t,\lambda}$, averaged over time and all wavelengths. For reconstructions with respectively $n_e$=1 up to $n_e$=5 elementary spectra, the normalised mean square error equals 3.5\%, 0.36\%, 0.21\%, 0.18\%, and 0.13\%. With one single spectrum, the decomposition is trivial. Two spectra clearly improve the quality of the fit. Some improvement is still noticeable with $n_e$=3 spectra, but the error then levels off because the model starts fitting noise. Thus, according to the error criterion $J$, the number of elementary spectra should be two or three.

The number of spectra can also be determined by inspection. With two sources, one elementary spectrum reproduces the quiet Sun, and the other captures a blend of coronal and transition region lines. With three sources, as we shall see below, the sources clearly separate lines that are generated at different temperatures. With four sources, one of the spectra becomes degenerated, as it appears twice with almost the same content\footnote{The difference between the two spectra captures a small instrumental perturbation  caused by temperature variations in the spacecraft. This effect and solutions for mitigating it with the BPSS will be discussed in a forthcoming paper.}. We conclude that the spectral variability between 25 and 195 nm is best described by the superposition of three elementary spectra. From now on, we stick to these three spectra.

\begin{figure}
\centering
\includegraphics[width=0.98\columnwidth]{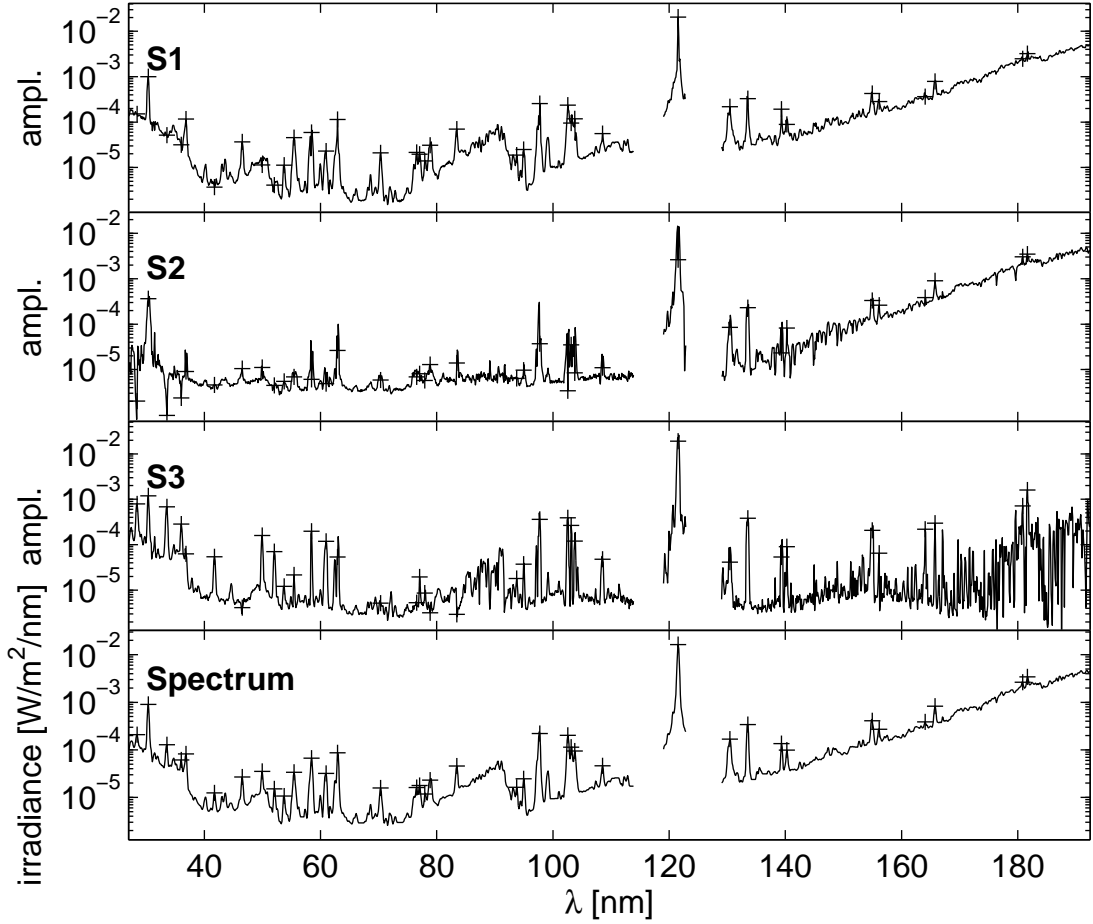}
\caption{The three elementary spectra (S1-S3, in arbitrary units) and the solar spectrum, all averaged over the whole time span. Crosses denote 38 intense spectral lines.}
\label{fig:spectra}
\end{figure}

\begin{figure}
\centering
\includegraphics[width=0.98\columnwidth]{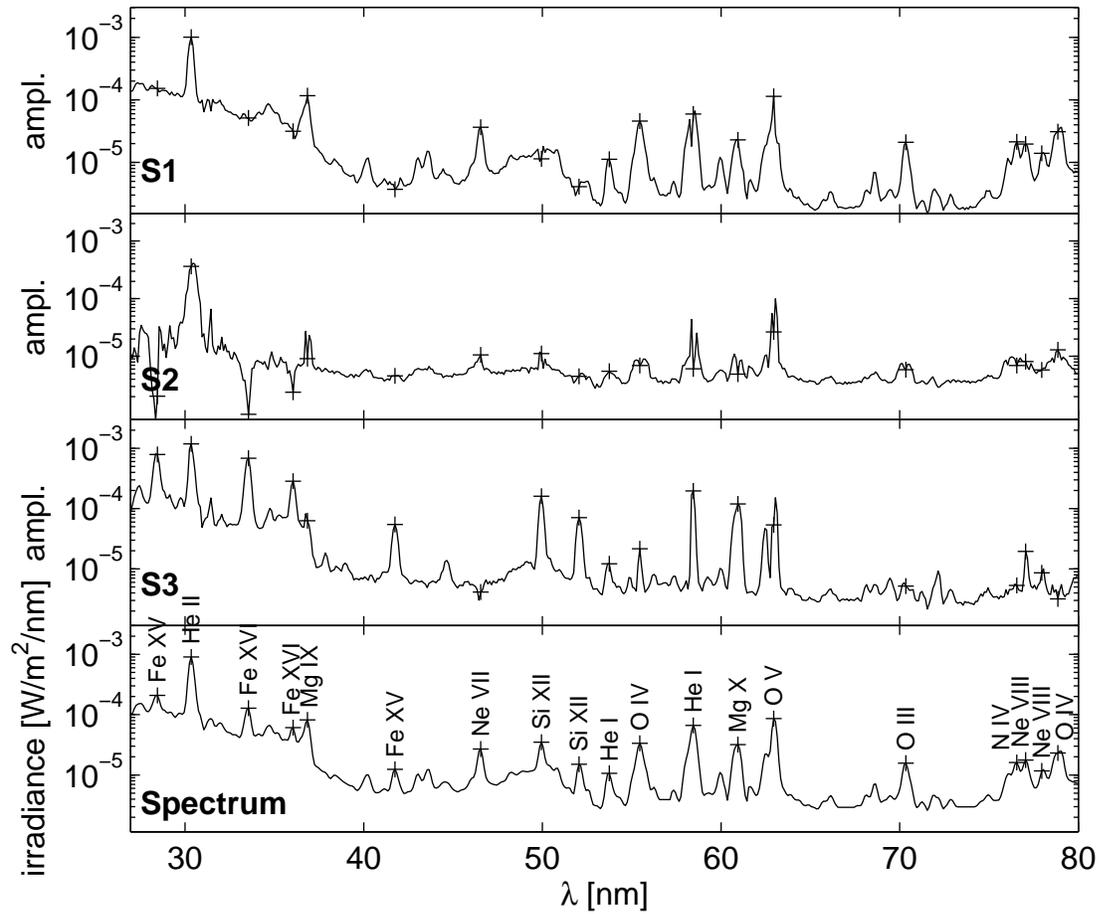}
\caption{Excerpt of Fig.~\ref{fig:spectra}, showing the detail of the elementary spectra between 28 and 80 nm, as well as the average solar spectrum.}
\label{fig:excerpt}
\end{figure}

The three elementary spectra $S_i(\lambda)$, estimated using BPSS, are shown in Fig.~\ref{fig:spectra}, with an excerpt in Fig.~\ref{fig:excerpt}. The first elementary spectrum (S1) looks similar to the time-averaged EUV spectrum, but it is not: S1 reproduces most of the strongest contributions, such as the intense HI Lyman-$\alpha$ line (121.57 nm), and the thermal continuum above 130 nm. Figure~\ref{fig:excerpt}, however, shows that hot coronal lines such as Fe XV (28.45 and 41.75 nm) and Fe XVI (33.55 and 36.05 nm) are   significantly reduced or even totally lacking, while transition region lines such as Ne VII (46.55 nm) and, for instance, the HI Lyman continuum are still present. Since a  major fraction of S1 comes from the  cooler part of the solar atmosphere, where most of the EUV radiation originates, we interpret it as  an average inactive Sun, defined as a full Sun, without important signs of activity.

%%% figure avec les spectres
%%%

%%% figure avec extrait des spectres
%%%

The second elementary spectrum S2 is more enigmatic, since it captures the thermal continuum, but not very many spectral lines apart from chromospheric ones such, as Si II (180.85 and 181.65 nm) and the wings of He II (30.35 nm). As we see later, this second spectrum mostly captures the contribution  from the coolest part of the chromosphere. 

The third elementary spectrum S3 stands out by the absence of the thermal continuum and the marked presence of hot coronal lines, such as Fe XVI (33.55 and 36.05 nm) and also Si XII (49.95 and 52.05 nm). Note the absence of the thermal background contribution and how the wings of blended lines are rejected. The third spectrum can therefore be interpreted as a contribution from hot coronal emissions.

\begin{figure}
\centering
\includegraphics[width=0.98\columnwidth]{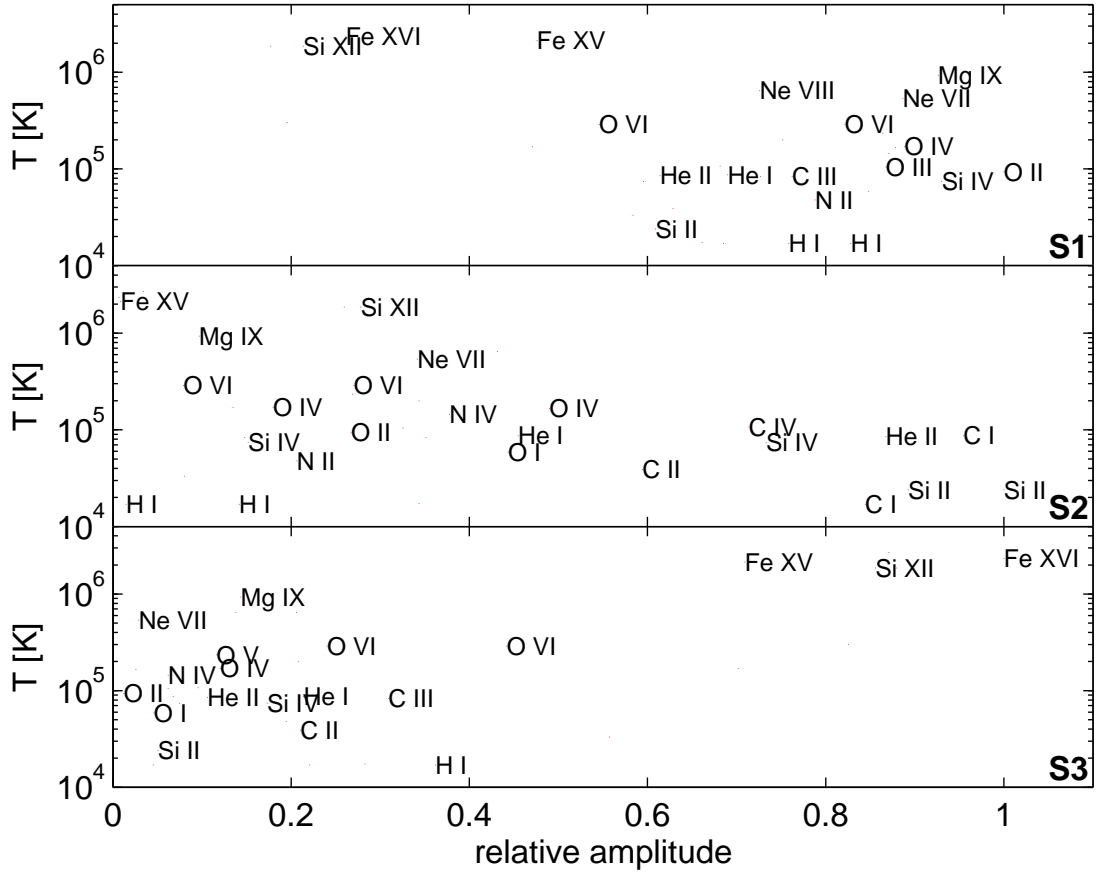}
\caption{For each of the 38 intense spectral lines shown in Fig.~\ref{fig:spectra}: characteristic emission temperature versus their relative contribution to each of the three elementary spectra. Some lines have been omitted to avoid excessive cluttering. The temperature is estimated using the CHIANTI model, and the characteristics of the SEE spectrometer. Lines with a bimodal temperature response are not shown on this plot.}
\label{fig:temperature}
\end{figure}

Clearly, our three elementary spectra do not correspond to specific regions of the Sun, and so cannot be directly compared to reference spectra as obtained from single instruments. To put our interpretations on firmer ground, we consider the effective temperature of 38 intense lines in Fig.~\ref{fig:temperature}. For each of them, we plot the contribution to the three elementary spectra, relative to the measured average spectrum, versus the effective temperature. The latter takes the finite spectral resolution of TIMED/SEE into account for blended lines.  

%%% amplitude envers la temperature
%%%

Figure~\ref{fig:temperature} confirms the temperature ordering of the elementary sources, as discussed above. The first elementary spectrum emphasises chromospheric and transition region lines, while S2 selects only the coolest chromospheric lines and S3 hot coronal lines, so from statistical properties only of the spectral variability, we can decompose the solar spectral irradiance into a unique set of three elementary contributions that correspond to specific temperature bands. To the best of our knowledge, this is the first proof of the existence of such a decomposition by rigourous means.

\begin{figure}[t!]
\centering
\includegraphics[width=0.98\columnwidth]{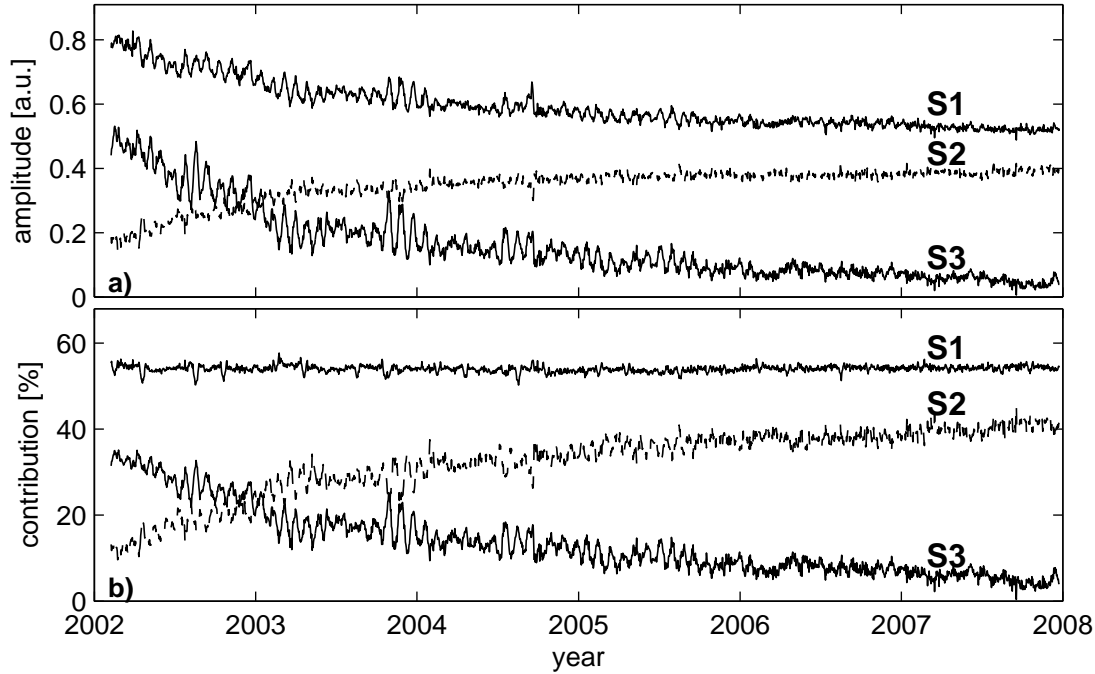}
\caption{Upper plot: absolute contribution of each source to the irradiance. Lower plot: Relative contribution of each source to the irradiance. The three terms add up to 100\%.}
\label{fig:concentration}
\end{figure}

Let us now investigate how the contributions of the elementary spectra evolve in time. Figure~\ref{fig:concentration} shows  the absolute contribution $V_k$, as well as the relative contribution $V_k/(V_1+V_2+V_3)$ for each spectrum. The data set covers the declining phase of the solar cycle, during which the irradiance drops at all wavelengths. As expected, the relative contribution of S3 gradually decreases in time, as active regions become scarce. The relative amplitude of S1 stays remarkably constant. This was to be expected,  as S1 is dominated by the contribution from the dominant Lyman-$\alpha$ line (121.57 nm) and the thermal continuum. As a consequence  of this, source S2 increases, both in relative and in absolute terms. This increase points toward a stronger contribution of the cold chromosphere at solar minimum.

To validate the interpretation of the elementary spectra, we compare their intensity $V_k(t)$ to various proxies for solar activity, using the cosine distance
$$
\gamma_{xy} = \langle x(t) y(t) \rangle \big/ \left(   \sqrt{\langle x^2(t) \rangle} \sqrt{\langle y^2(t) \rangle} \right)
$$
rather than the usual correlation coefficient. Both quantities have the same interpretation and are bounded by $[-1,1]$; the cosine distance, however, takes into account the magnitude of the relative variability during the solar cycle. Although none of the indices can satisfactorily reproduce the spectral variability on both short and long time scales \citep{ddw08}, high values of $\gamma_{xy}$ should nevertheless hint at the origin of the spectra. 

The two indices that are most strongly correlated with the first source are by far the MgII core-to-wing \citep{viereck01} and the CaII K \citep{lean82} indices, with $\gamma_{xy} = 0.995$ and $\gamma_{xy} = 0.996$, respectively. Both indices are indeed known to reproduce the UV spectrum well. The second source does fit no known index, since it increases over the declining cycle. The third source, however, is strongly correlated with the Mount Wilson Sunspot Index \citep{parker98}  and to a lesser degree with the radiometric f10.7  index \citep{tobiska00}, with  $\gamma_{xy}=0.98$ and $\gamma_{xy}=0.95$, respectively.  Both indices quantify only the contribution from active regions. These results therefore fully support our interpretation of the first and the third spectra. They also suggest the possibility of reconstructing the spectrum and its variability using properly chosen proxies. This should be straightforward for sources S1 and S3, whereas more work is needed to model source S2. We are currently working on this problem.

%%% figure avec contributions des 3 sources
%%%
 
% \vspace{-.2cm} 
\section{Discussion and outlooks}

The central result of this study is the possibility to describe the EUV spectral variability in terms of only three spectra. This confirms (and in some sense justifies) a long standing and purely intuitive practice that consisted in partitioning the Sun into three components. Our approach provides a new interpretation of these components, excluding any a priori bias. The prevalent viewpoint assumed a horizontal structuring of the solar emission in terms of coronal holes, active regions, active network, etc. Our results instead support the existence of emitting volumes, with a more vertical structuring. The difference is important, as it raises the questions of the energy exchange between the lower atmosphere and the corona and of its evolution with the solar cycle. This interpretation will be developed in a forthcoming paper.

Our results also pave the way for a new instrumental concept in which the solar EUV spectrum would be reconstructed only from a small set of lines or spectral bands, rather than using a full-fledged spectrometer. The reason for this is the remarkable redundancy of the spectral variability, which has already been revealed by \citet{ddw05}, using multivariate statistical analysis. Such a redundancy necessarily implies a strong connection between the physical processes at different solar atmospheric layers.  

Another interesting result is the behaviour of the second and third elementary spectra in Fig.~\ref{fig:concentration}, which supports a gradual migration of the origin of EUV flux from the low corona and high transition region to the low transition region and high chromosphere. In a forthcoming study, we shall determine how the heat flux in the transition region could be forced by this behaviour along the solar cycle. But first, it should be confirmed during the decreasing part of the solar cycle. And this leads to the following issue: does the minimum of S2 coincide with solar maximum, if any, or could it be the crossing between S2 and S3? We probably will need more than a solar cycle of observations to answer this question.

%%%%%%%%%%%%%%%%%%%%%%%%%%%%%%%%%%%%%%%%%%%%%%%%%%%%%
% \vspace{-.2cm}
\subsection*{Acknowledgements}
We wish to thank the TIMED/SEE team for making the spectral irradiance data available and Tom Woods for valuable comments that substantially improved the manuscript. This work was supported by the French \emph{Programme National Soleil-Terre}.

%%%%%%%%%%%%%%%%%%%%%%%%%%%%%%%%%%%%%%%%%%%%%%%%%%%%%
%... bibliography
% \vspace{-.3cm}
\bibliographystyle{aa}

\begin{thebibliography}{19}
\expandafter\ifx\csname natexlab\endcsname\relax\def\natexlab#1{#1}\fi

\bibitem[{{Curdt} {et~al.}(2001){Curdt}, {Brekke}, {Feldman}, {Wilhelm},
  {Dwivedi}, {Sch{\"u}hle}, \& {Lemaire}}]{curdt01}
{Curdt}, W., {Brekke}, P., {Feldman}, U., {et~al.} 2001, A\&A, 375,
  591

%\bibitem[{{Dudok de Wit} {et~al.}(2008){Dudok de Wit}, {Kretzschmar},
%  {Aboudarham}, {Amblard}, {Auch\`ere}, \& {Lilensten}}]{ddw08}
%{Dudok de Wit}, T., {Kretzschmar}, M., {Aboudarham}, J., {et~al.} 2008, Adv.
%  Space Res., in press

\bibitem[{{Dudok de Wit} {et~al.}(2008){Dudok de Wit}, {Kretzschmar},
  {Aboudarham}, {Amblard}, {Auch\`ere}, \& {Lilensten}}]{ddw08}
{Dudok de Wit}, T., {Kretzschmar},  {et~al.} 2008, Adv.
  Space Res., in press



%\bibitem[{{Dudok de Wit} {et~al.}(2005){Dudok de Wit}, {Lilensten},
%  {Aboudarham}, {Amblard}, \& {Kretzschmar}}]{ddw05}
%{Dudok de Wit}, T., {Lilensten}, J., {Aboudarham}, J., {Amblard}, P.-O., \&
%  {Kretzschmar}, M. 2005, Annales Geophysicae, 23, 3055

\bibitem[{{Dudok de Wit} {et~al.}(2005){Dudok de Wit}, {Lilensten},
  {Aboudarham}, {Amblard}, \& {Kretzschmar}}]{ddw05}
{Dudok de Wit}, T., {Lilensten}, {et~al.} 2005, Annales Geophysicae, 23, 3055

\bibitem[{{Floyd} {et~al.}(2005){Floyd}, {Newmark}, {Cook}, {Herring}, \&
  {McMullin}}]{floyd05}
{Floyd}, L., {Newmark}, J., et al. 2005, J. Atmos. Terr. Phys., 67, 3

\bibitem[{{Fontenla} \& {Harder}(2005)}]{fontenla05}
{Fontenla}, J. \& {Harder}, G. 2005, Mem. della Soc. Astron.
  Italiana, 76, 826

\bibitem[{Gelman {et~al.}(2003)Gelman, Carlin, Stern, \& Rubin}]{gelman03}
Gelman, A., {et~al.}  2003, Bayesian Data
  Analysis (Chapman and Hall)

\bibitem[{{Kretzschmar} {et~al.}(2004){Kretzschmar}, {Lilensten}, \&
  {Aboudarham}}]{kretzschmar04}
{Kretzschmar}, M., {Lilensten}, J., \& {Aboudarham}, J. 2004, A\& A, 419, 345

\bibitem[{{Kretzschmar} {et~al.}(2006){Kretzschmar}, {Lilensten}, \&
  {Aboudarham}}]{kretzschmar06}
{Kretzschmar}, M., {Lilensten}, J., \& {Aboudarham}, J. 2006, Adv. Sp.
  Res., 37, 341

%\bibitem[{{Lean} {et~al.}(1982){Lean}, {Livingston}, {Heath}, {Donnelly},
%  {Skumanich}, \& {White}}]{lean82}
%{Lean}, J.~L., {Livingston}, W.~C., {Heath}, D.~F., {et~al.} 1982, J. Geophys.
%  Res, 87, 10307

\bibitem[{{Lean} {et~al.}(1982){Lean}, {Livingston}, {Heath}, {Donnelly},
  {Skumanich}, \& {White}}]{lean82}
{Lean}, J.~L., {Livingston}, W.~C.,   {et~al.} 1982, J. Geophys.
  Res, 87, 10307

%\bibitem[{Moussaoui {et~al.}(2006)Moussaoui, Brie, {Mohammad-Djafari}, \&
%  Carteret}]{moussaoui06}
%Moussaoui, S., Brie, D., {Mohammad-Djafari}, A., \& Carteret, C. 2006, IEEE
%  Trans. Signal Processing, 11, 4133

\bibitem[{Moussaoui {et~al.}(2006)Moussaoui, Brie, {Mohammad-Djafari}, \&
  Carteret}]{moussaoui06}
Moussaoui, S., Brie, D., {et~al.}   2006, IEEE
  Trans. Sig. Proc., 11, 4133

\bibitem[{Paatero \& Tapper(1994)}]{paatero94}
Paatero, P. \& Tapper, U. 1994, Environmetrics, 5, 111

\bibitem[{{Parker} {et~al.}(1998){Parker}, {Ulrich}, \& {Pap}}]{parker98}
{Parker}, D.~G., {Ulrich}, R.~K., \& {Pap}, J.~M. 1998, Solar Physics, 177, 229

\bibitem[{{Richards} {et~al.}(1994){Richards}, {Fennelly}, \&
  {Torr}}]{richards94}
{Richards}, P.~G., {Fennelly}, J.~A., \& {Torr}, D.~G. 1994, J. Geophys. Res.,
  99, 8981

%\bibitem[{{Tobiska} {et~al.}(2000){Tobiska}, {Woods}, {Eparvier}, {Viereck},
%  {Floyd}, {Bouwer}, {Rottman}, \& {White}}]{tobiska00}
%{Tobiska}, W.~K., {Woods}, T., {Eparvier}, F., {et~al.} 2000, Journal of
%  Atmospheric and Terrestrial Physics, 62, 1233

\bibitem[{{Tobiska} {et~al.}(2000){Tobiska}, {Woods}, {Eparvier}, {Viereck},
  {Floyd}, {Bouwer}, {Rottman}, \& {White}}]{tobiska00}
{Tobiska}, W.~K., {Woods} T.N. , {et~al.} 2000, J.
  Atmos. Terr. Phys., 62, 1233


\bibitem[{{Vernazza} \& {Reeves}(1978)}]{vernazza78}
{Vernazza}, J.~E. \& {Reeves}, E.~M. 1978, Astroph. J. Supp. Ser., 37, 485


% \bibitem[{{Vernazza} {et~al.}(1981){Avrett}, {Loeser}}]{vernazza81}
% Vernazza,  {et~al.} 1981, Astrophys. J.  
% Supp. Ser 45, 635


\bibitem[{{Viereck} {et~al.}(2001){Viereck}, {Puga}, {McMullin}, {Judge},
  {Weber}, \& {Tobiska}}]{viereck01}
{Viereck}, R., {Puga}, L., {McMullin}, D., {et~al.} 2001, Geoph. Res. Lett.,
  28, 1343

\bibitem[{{Warren}(2005)}]{warren05}
{Warren}, H.~P. 2005, Astroph. J. Suppl. Ser., 157, 147

\bibitem[{{Warren} {et~al.}(2001){Warren}, {Mariska}, \& {Lean}}]{warren01}
{Warren}, H.~P., {Mariska}, J.~T., \& {Lean}, J. 2001, J. Geophys. Res., 106,
  15745
  
  
  \bibitem[{{Wenzler} {et~al.}(2006)}]{wenzler06}
{Wenzler}, T. , {et~al.} 2006, A  \& A , 460,  583 
%\bibitem[{{Woods} {et~al.}(2005){Woods}, {Eparvier}, {Bailey}, {Chamberlin},
%  {Lean}, {Rottman}, {Solomon}, {Tobiska}, \& {Woodraska}}]{woods05}
%{Woods}, T.~N., {Eparvier}, F.~G., {Bailey}, S.~M., {et~al.} 2005, J. Geophys.
%  Res., 110, 1312

\bibitem[{{Woods} {et~al.}(2005){Woods}, {Eparvier}, {Bailey}, {Chamberlin},
  {Lean}, {Rottman}, {Solomon}, {Tobiska}, \& {Woodraska}}]{woods05}
{Woods}, T.~N., {Eparvier}, F.,  {et~al.} 2005, J. Geophys.
  Res., 110, 1312

\bibitem[{{Woods} {et~al.}(2000){Woods}, {Tobiska}, {Rottman}, \&
  {Worden}}]{woods00}
{Woods}, T.~N., {Tobiska}, {et~al.} J.~R. 2000, J.
  Geophys. Res, 105, 27195

\end{thebibliography}

\end{document}